\documentstyle[12pt,prl,aps,epsfig]{revtex}

\begin{document}
\draft
\title{Universal Cubic Eigenvalue Repulsion for Random Normal 
Matrices}
\author{Gary Oas\cite{GO}}
\address{Ventura Hall, Stanford University, Stanford, CA 94305 \\
{\it To appear in Physical Review E}}

\date{ \today}
\maketitle
\begin{abstract}
Random matrix models consisting of normal matrices, 
defined by the sole constraint $[{\bf N} ^{\dag},{\bf N} ]=0$, will be
explored.
It is shown that cubic eigenvalue repulsion in the complex
plane is universal with respect to the probability distribution
of matrices.
The density of eigenvalues, all correlation functions, and level
spacing statistics are calculated.
Normal matrix models offer more probability distributions 
amenable to analytical analysis than complex matrix models
where only a model with a Gaussian distribution are solvable.
The statistics of numerically generated eigenvalues from gaussian
distributed normal matrices are compared to the analytic results
obtained and agreement is seen.
\end{abstract}



\section{Introduction}

Random matrix theory (RMT), \cite{Wigner}, \cite{mehta} has found much
success as phenomenological
models describing a wide variety of physical systems,
from discretization
of moduli space \cite{kontsevich}, to the statistics of the cells in the skin of a cucumber
\cite{voronoi}.
RMT is the study of eigenvalues derived from
random ensembles of matrices with stochastic elements specified by a 
probability density, $P({\bf M} )d{\bf M} $, in the space of matrices.
Most interest is in examining the properties of the eigenvalues induced
from the transformation to the eigenvalue basis. Early work
consisted of using real symmetric, hermitian, unitary, and 
real quarternion matrices \cite{Dyson}. The eigenvalues of which are either
real or unimodular, (and termed one dimensional eigenvalues here).

Recently several groups have begun to consider physical applications
of matrix models composed of complex matrices. 
Introduced in the early sixties by Ginibre \cite{ginibre}
it took decades for others to consider applications. 
As the probability distribution
is not invariant under similarity transformations
the diagonalizing parameters must be integrated out by brute force.
It was found that only
a distribution with a gaussian weight (Ginibre Ensemble) 
could be solved.
The Ginibre ensemble was later found to exhibit cubic eigenvalue 
repulsion in the complex plane, \cite{haake} \cite{mehta}.
Such models are of interest in characterizations of quantum chaos.
Cubic quasi-energy level repulsion is a key signature of classical
chaos within a quantum dissipative system, as defined by Haake \cite{haake}.

Normal matrices are discussed in most matrix theory
texts, (see \cite{matext}, \cite{macklin}).
  Defined by the sole constraint that they commute with
their adjoint, they have the property of being the most
general matrix that can be diagonalized by a unitary transformation.
The normal matrix model was first introduced in showing how the
Laughlin wavefunction could be modeled by it and offered some
generalizations to inhomogeneous fields \cite{chau}.
In \cite{oas} the statistics of eigenvalues of random normal
matrices were first explored.

In this paper 
the universality of cubic eigenvalue
repulsion in the complex plane for ensembles of random normal
matrices will be shown. 
In section II the probability distribution in the space of matrices,
all correlation functions, and the eigenvalue densities will be
obtained.
In section III the level spacing statistics of the eigenvalues obtained will be
derived. The term level spacing is borrowed from the Wigner-Dyson
ensembles and will represent here the spacing in the complex plane.
In section IV eigenvalues of numerical generated normal matrices will be 
studied and compared to the analytical results.

\section{Probability Distributions, N-pt. Correlation Functions}

We begin by defining the joint probability distribution (jpd) within the space of
normal matrices,
\begin{eqnarray}
P_N ({\bf M} ) &=& C \exp [-Tr( V({\bf M} ,{\bf M}^{\dag} ))].
\end{eqnarray}
Only potentials which are hermitian will be considered, $V({\bf M} ,{\bf M}^{\dag})
=V({\bf M} ,{\bf M}^{\dag})^{\dag} $.
As a result the joint probability distribution
for the eigenvalues will be rotationally symmetric in the complex plane.
Normal matrices are the most general matrices which can be diagonalized 
by a unitary transformation,
 $\bf{M} = \bf{ UZU^{\dag}} \label{untrans}$,
so that
the measure and the weight are invariant.
Thus it is simple to derive the 
jacobian for the transformation to the eigenvalues.

Proceeding analogously to the hermitian matrix case \cite{Tracy1},  
the metric in the space of normal matrices is defined as,
\widetext

\begin{eqnarray}
(ds)^2 & = & Tr[\bf{dM^{\dag}dM}] \nonumber \\ 
& = &  Tr(\bf{Ud(U^{\dag}ZU)U^{\dag}Ud(U^{\dag}Z^{\ast}U)U^{\dag}})\nonumber \\
& = & Tr(\bf{[dS,Z][dS,Z^{\ast}]}+(\bf{[dS,Z]dZ^{\ast}} + \bf{ dZ[dS,Z^{\ast}]} ) + \vert \bf{dZ}\vert^2) \nonumber \\ 
&  = & c \sum_{i<k}\vert dS_{ik}(z_k - z_i)\vert^2 + \sum_i \vert dz_i\vert^2 ,
\end{eqnarray}
where $\bf{dS}=\bf{UdU^{\dag}}=-\bf{dU U^{\dag}}$ is antihermitian
and $c$ is an overall constant.

Using the standard Riemannian volume form with this metric we get,

\begin{eqnarray}
d\mu ({\bf M}) \longrightarrow d\mu(z) = (d\Omega) \vert\Delta (z)\vert^2 dz^{(0)} _1 \wedge dz^{(0)} _2 \wedge ... \wedge dz^{(0)} _N \wedge dz^{(1)} _1 \wedge ...\wedge dz^{(1)} _N . 
\end{eqnarray}
where $d\Omega $ is the volume of the unitary group $U(N)$, and 
$\Delta (z)$ is the well known Van der Monde determinant, 

\begin{eqnarray}
|\Delta (z)|^2 &=& \prod_{i<j}^{N}|z_j-z_i|^2
\end{eqnarray}


If we consider weights
which are invariant under unitary transformations 
we can factor $d\Omega $ out,

\begin{eqnarray}
P_N(z) &=& C' |\Delta (z)|^2 \exp [-\sum_{i=1}^N V(z_i,z_i ^{\ast})],
\label{probdis}
\end{eqnarray}
where $C'$ is a normalization constant.

Considering a gaussian weight, $V=|z|^2$, we notice that
the jpd is identical to Ginibre's ensemble of complex matrices\cite{ginibre}.
The model with this weight function
is termed the Normal Gaussian Ensemble (NGE).
It is clear that gaussian ensembles of complex matrices and the 
NGE are identical.
Any result derived here for the NGE will also be valid for Ginibre's 
ensemble of matrices. This is of little surprise since 
$Tr[{\bf M}^{\dag}{\bf M}]=Tr[{\bf M}{\bf M}^{\dag}]$.

The difference occurs when considering other ensembles. For complex
matrices the parameter $\beta$, (defined as the power of the Van der Monde
determinant appearing in the jpd), is a function of the weight.
In fact, only models
involving gaussian weights have ever been studied
as this is the only nontrivial model which allows a separation of 
radial and angular parameters. 
For normal matrices $\beta =2$
regardless of the weight and it is possible to study a wide variety of 
ensembles.

For matrix models with one dimensional eigenvalues 
 the traditional method of analysis
employs a basis of orthogonal polynomials. These allow the reduction
of the determinant upon integration of a number of eigenvalues \cite{mehta}.
For complex eigenvalues we can also introduce an orthogonal basis.
If a polar basis is chosen, (and $P_N (z)$ is rotationally symmetric), it is simple 
to verify that the $z$'s themselves form an orthogonal basis.  
 A basis of {\it orthogonal monomials}, $\{ \phi \} $, is defined,

\begin{eqnarray}
\phi _l (z_i) &\equiv& {z_i ^l\over N _l ^{1\over 2}},
\end{eqnarray}
where $N_l=\int_0^{\infty}dr^2 r^{2l}\exp [-V(r)]$, is a normalization
constant.

The jpd can then be expressed as,
\begin{eqnarray}
P_N (z) &=& C' e^{-\sum_{i=1}^N V(z_i,z_i ^{\ast} )}det K_N (z_i,z_j)|_{i,j=1,N},
\end{eqnarray}
where the ``kernel'' is,
\begin{eqnarray}
K_N(z_i,z_j)&=& \sum_{l=0}^{N-1} \phi_l (z_i) \phi_l ^{\ast} (z_j).
\end{eqnarray}
The weight function is deliberately factored out of the kernel since all 
of the angular dependence is contained within $K$.

Similarly to hermitian matrix models, when $N-n$ eigenvalues are integrated
out of the jpd, the determinant shrinks by $N-n$ columns and rows, \cite{oas}.
This property allows the $n$-point correlation function to be
derived easily,
\begin{eqnarray}
R_n(z_1,\cdots ,z_n)&\equiv& {N!\over (N-n)!}\int \prod_{i=n+1}^N 
{1\over 2}d\theta _i dr_i ^2 P_N(z), \nonumber \\
&=& e^{-\sum_{i=1}^n V(r_i)} det K_N(z_i,z_j)|_{i,j=1,n}.
\end{eqnarray}
The prefactor ${N!\over (N-n)!}$ is due to the ordering of eigenvalues.
The 2 point correlation function and eigenvalue density
are,
\begin{eqnarray}
R_2 (z,w) &=&{
 e^{-V(z)-V(w)}\over \pi ^2}[\rho (z)\rho (w)-K(z,w)K(w,z)],  \nonumber  \\
\rho (z) &=& e^{-V(z)} \sum_{l=0}^{N-1} {|z|^2\over N_l},
\end{eqnarray}
In deriving expressions for the correlation functions
orthogonal monomials provide a more economical method.
However, there are no simple recursion relations among
the monomials as there are for the orthogonal polynomials.    
No recursion relation implies no Christoffel-Darboux formula and no
asymptotic analysis. 
However, we can find 
analytical results for finite $N$. 
It is not necessary to examine the asymptotic form of the monomials to
get large $N$ forms.

In table \ref{table} various ensembles are defined and their
corresponding eigenvalue density and two point correlation function
are given.
It should be noticed that several of these ensembles allow {\it global}
closed forms in the limit $N\rightarrow\infty$.

\section{Level Spacing Statistics}

In order to study the properties of random matrices the statistical
properties of the spacings between eigenvalues, 
(or level spacings), are examined.
These statistics can be used as a definition of quantum chaos.
The distinction of classically integrable and chaotic systems
can be seen in a semiclassically quantized system as the transition
from Poisson to Wigner-Dyson statistics of the nearest neighbor
distribution of the energy levels \cite{berry}.
It has been shown \cite{haake} that
if there is dissipation in the system the level spacings undergo
a similar transition from a Plane Poisson distribution, (random
points in a plane), to Ginibre's distribution \cite{ginibre} generated by an 
ensemble of random complex matrices. 

The distributions for eigenvalues in the complex plane will require
the definition of statistical quantities analagous to those defined
for Hermitian matrix models \cite{mehta}.
\begin{itemize}
\item The probability that no eigenvalue lies within a circle of radius $s$
centered upon the point $w$, the {\it gap distribution}, 
is denoted as $E(s,w)$. A related quantity $E(s,w,\epsilon )$ is defined
as the probability that no eigenvalues lie within an annulus of outer
radius $s$, inner radius $\epsilon$, and centered on the point $w$.

\item The probability that no eigenvalues lie within this same circle but 
one or more eigenvalues lying on the perimeter is denoted as $F(s,w)$.

\item For an eigenvalue lying at the center of the circle and no eigenvalues
within the circle nor on the edge, this probability is denoted by $H(s,w)$.

\item The level spacing probability, $p(s,w)$, is the most quoted quantity and 
is defined as the probability that one eigenvalue lies at the center and one ormore on the perimeter.
\end{itemize}

By taking differential areas at the center and at the perimeter of this 
circle, we can obtain all of the probabilities from the gap distribution.
The argument is analagous to Mehta's for one dimensional eigenvalues 
and can be found in \cite{oas}.

\begin{eqnarray}
E(s,w)&=&\int \{ dz \} P_N(\{ z \} ) \prod_{i=1}^N [1- \chi (z_i,s,w)], \label{esw}\\
F(s,w)&=& -{d\over ds} E(s,w) \\
H(s,w)&=& -{d\over d\epsilon ^2 } E(s,w,\epsilon )|_{\epsilon =0} \\ 
p(s,w)&=& -{d\over ds}H(s,w) =\label{psw} \\
&&  -{d\over d\epsilon ^2 }F(s,w,\epsilon )|_{\epsilon =0} = {d\over ds}{d\over
d\epsilon }E(s,w,\epsilon )|_{\epsilon =0}. \nonumber 
\end{eqnarray}
Where, $\chi$ is the characteristic function and $\epsilon$ is taken as the radius
of an infinitesmal area centered at the center of the circle.  
These relations allow an elegant derivation of the level spacing
distribution.

All of the probability distributions considered here are symmetric in
the $z_i$, hence the derivation of the gap distribution will be the
same as in \cite{mehta}, (see also \cite{oas}). 
For the circle centered
at the origin we have the following,
\begin{eqnarray}
E(s,0) &=& \prod_{i=0}^{N-1} \left[ {\int_{s^2}^{\infty} dx \exp (-V(x))x^j\over
\int_0^{\infty} dx \exp (-V(x))x^j}\right], \label{gap}
\end{eqnarray}
where $x= |z|^2$. 

As a check on our results, we find the spacing distribution for the
NGE and compare with previously known results.
In this case the probability distribution is translationally invariant, 
hence
the result (\ref{gap}) is valid for all $w$. The result for small $s$
is,

\begin{eqnarray}
p(s) \approx 2s^3-s^5+{1\over 7} s^7-{11\over 12}s^9 \cdots \label{ngueps}
\end{eqnarray}
This result is the same as for complex matrices derived by Haake\cite{haake}.
The level spacing statistics at the origin for other ensembles of
normal matrices can be found in \cite{oas}. Here we are concerned with
obtaining a universal result for the spacing distribution.

For ensembles other than Gaussian, the distributions are not translationally
 invariant and hence
it will be necessary to resort to asymptotic approximations.
Shifting the eigenvalue coordinate, $z=z-w$, does not affect the measure
or Van der Monde determinant. The gap distribution can then be
expressed as, 
\widetext
\begin{eqnarray}
E(s,w) &=& \prod_{i=1}^N \int_0^{\infty} dr_i ^2 d\theta _i
 [1-\chi_i (s^2)] {\tilde P} _N( z), \label{esw1}\nonumber \\
&=& \int_0^{\infty}\{ d^2z\} \{1-\sum_{i=1}^N \chi_i (s^2) +\sum_{i\neq j = 1}^N
\chi _i (s^2)\chi _j (s^2)-\cdots \}{\tilde P}_N(z), \nonumber \\
&=& 1-\sum_i^N \int d^2z_i \chi_i \tilde\rho (z_i) +\sum_{i\leq j}^N \int d^2z_i d^2z_j
\chi _i \chi _j {\tilde R} _2 (z_i ,z_j)-\cdots , \label{es2}
\end{eqnarray}
where $\tilde f(z) = f(z-w)$ and $\chi _i$ is the 
characteristic function for the $i^{th}$ eigenvalue.
In the limit $s\rightarrow 0$, and keeping only the first two terms
we obtain,
\begin{eqnarray}
\lim_{ s\rightarrow 0}E(s,w)&\approx & 1-N \pi s^2 \rho (w).
\end{eqnarray}

The level spacing distribution, $p(s,w)$, can be found via equation 
(\ref{psw}). First the gap distribution $E(s,w,\epsilon )$ needs to be derived.
The derivation is the same as before except for a modification to the
characteristic function. Starting from (\ref{es2}) we have,
\widetext
\begin{eqnarray}
E(s,w,\epsilon ) &=& 1-\sum_i^N \int d^2z_i \chi_i \tilde\rho (z_i) +\sum_{i\leq j}^N \int d^2z_i d^2z_j
\chi _i \chi _j {\tilde R} _2 (z_i ,z_j)-\cdots ,
\end{eqnarray}
where now $\chi_i = \theta (s^2-r_i ^2) \theta (r_i ^2 - \epsilon ^2)$.
As an intermediate step we find,

\begin{eqnarray}
H(s,w) &=& -{d\over d\epsilon ^2}E(s,w,\epsilon )|_{\epsilon =0}, \nonumber \\
&=& N {d\over d\epsilon^2 } \int_{\epsilon ^2}^{s^2}dr^2\int d\theta
\rho (z+w) 
\nonumber \\
&& \qquad -{N^2 +N\over 2} {d\over d\epsilon ^2} \int_{\epsilon ^2}^{s^2}\int_{\epsilon ^2}^{s^2}{dr_1 ^2 dr_2 ^2\over 4}
\int\int d\theta _1 d\theta _2  R_2 (z_1 +w,z_2+w) +\cdots |_{\epsilon=0},
 \nonumber \\
&=& -N\pi\rho (w)+\pi (N^2+N)\int_0^{s^2}{dr_2 ^2\over 2}\int d\theta _2
R_2 (w,z_2+w) -\cdots. 
\end{eqnarray}
The last term exploits the symmetry in $R_2$. 
For the level spacing we have,
\begin{eqnarray}
p(s,w) &=& -\pi (N^2+N) s \int d\theta _2 R_2 (w, se^{i\theta _2}+w)
\cdots \label{psw2}
\end{eqnarray}
In order to find the lowest order contribution to $p(s,w)$ the lowest
order of $R_2$ will need to be found.
For a model with weight $e^{-V(z)}$ we have,
\widetext
\begin{eqnarray}
R_2 (w,z+w) &=& e^{-V(w)}e^{-V(z+w)}\sum_{l=1}^N \sum_{m=0}^N[ {r_w^{2l}
(r^2+zw ^{\ast} +wz^{\ast} +r_w^2)^m\over N_lN_m}-{(zw^{\ast} +r_w^2)^l(wz^{\ast} +r_w^2)^m
\over N_lN_m}].  
\end{eqnarray}
The normalization
factor, $N_l$ is model dependent.
It is easily verified that terms with $l=0$ and $l=m$ vanish. All
higher order functions will have the same property.

The remaining angular integral over $d\theta_2$ in (\ref{psw2})
insures that no terms linear in $z$ or $z^{\ast}$ appear. 
It is also apparent that the leading term in $r_w$ cancels,
 (terms to order $z^0$). This yields the
lowest order of $R_2$ to be $s^2$. Combined with equation (\ref{psw}),
it is found that the nearest level spacing is cubic in $s$ as
$s\rightarrow 0$. 
A few exceptions occur when the weight function vanishes as $z$
vanishes. 
Specifically the Jacobi, Laguerre, ($a>1$) and generalized Gaussian ($a>1$)
will have a higher order repulsion. 
Our final result is thus:

{\it Normal matrix models with normalizable hermitian weight functions have
eigenvalues experiencing a minimum of cubic repulsion in the complex
plane.}

It is also worthwhile to examine the case of real normal matrices.
Here the constraint being that the matrix commutes with its transpose.
Such a matrix is diagonalizable by a unitary transformation and 
will in general have complex eigenvalues. Retracing the steps
above it is easy to see that the eigenvalues will also exhibit
a minimum of cubic repulsion in the complex plane.

\section{Numerical Results}

Whenever discussing random matrix theory it is always of interest to
include numerical simulation of large random matrices. These are
used to verify the analytical properties and supply support
for asymptotic forms, conjectures, etc.
When dealing with normal matrices however, we are in a
predicament. Consider the following problem:
\begin{itemize}
\item {\it Write down an arbitrary $2\times 2$ normal matrix that is not
Hermitian, Unitary, nor Real.}
\end{itemize}
It is not so simple. If one was asked to write down a hermitian matrix,
it could be done without thinking twice. For a normal matrix there  
is no simple parameterization.
It would most likely require the commutator to be calculated and a solution
found by trial and error.
Likewise, if you want a computer to generate normal matrices you 
must give it a parameterization. For a $2\times 2$ you 
can parameterize it as follows, given that 
${n_{11}}^{\ast} \neq {n_{22}}^{\ast}$, (remember there are $N^2+N=6$
independent real elements within a normal matrix),
\begin{eqnarray}
{\bf N} &=&\left( \matrix{n_{11} & n_{12} \cr
n_{12}^{\ast}({n_{11}-n_{22}\over n_{11}^{\ast}-n_{22}^{\ast}})
 & n_{22}\cr }\right).
\end{eqnarray}
 
 For arbitrary $N$ the constraint equations
become nonlinear and highly coupled. 
We will choose the following parameterization of normal matrices, 
define the independent elements as $n_{ij}$, (those along the diagonal
and above the diagonal). Define the elements that are constrained as
$f_{ji}$, (those which lie below the diagonal). For the $ji^{th}$
element the constraint equation is, 
\begin{eqnarray}
f_{ji} &=& {1\over (n_{jj}^{\ast}-n_{ii}^{\ast})}\left[n_{ij}^{\ast} (n_{jj}-n_{ii}) 
+ \sum_{k=1}^{i-1} n_{kj}^{\ast} n_{ki}
-  \sum_{m=j+1}^{N} n_{im}^{\ast} n_{jm}\right] 
\nonumber \\
&&-{1\over (n_{jj}^{\ast} -n_{ii}^{\ast})}
\left[ \sum_{k=1}^{i-1}f_{jk}f_{ik}^{\ast} - \sum_{m=j+1}^N f_{mi}f_{mj}^{\ast}
 - \sum_{l=i+1}^{j-1} (n_{lj}^{\ast} f_{li}-n_{il}^{\ast} f_{jl})\right]. \label{constraint}
\end{eqnarray}

Notice that the term in the first bracket contains only independent elements.
With this parameterization the probability of obtaining 
normal matrices which are hermitian, unitary, orthogonal, or 
real symmetric matrices becomes infinitesimal for large $N$. 
The subspaces for these types of matrices within the space of normal
matrices are of much lower dimensionality.
There are many solutions to this nonlinear equation. 
Another method must be devised to obtain a normal matrix.
It is also found that due to the nonlinearity, only normal gaussian
matrices can be generated. (Others not respecting the hermiticity and
normalizibility could be analyzed but is not of interest here).

Instead of directly attempting to produce random normal matrices we begin
with a random complex matrix and use an optimization algorithm to obtain a 
normal matrix. The following algorithm is the simplest to
implement but by no means the most efficient.

\begin{enumerate}
\item Parameterize the normal matrix as before: \\
Elements along the diagonal and above the diagonal are independent
parameters. Elements below the diagonal will be optimized.

\item Using the Box-Muller technique, generate random gaussian variables
and construct a $200\times 200$ complex gaussian matrix. This is the
initial condition.

\item Calculate the commutator and define the following error function,
$ tr[{\bf N,N^{\dagger}}]^2 = E$. The error function will be minimized below a
choosen threshold, (a value of 0.2 is used). 

\item Use gradient descent to adjust the constrained elements. If the
value of $E$ is below the choosen threshold then pass the matrix to the 
diagonalization routine. If $E$ hasn't converged after 500 iterations,
then reset the matrix and go to step 2.

\end{enumerate}

The results of a preliminary numerical study based on this algorithm
are now presented.
A full discussion of the algorithm used is to be found 
elsewhere, \cite{oas}. Due to limited computational resources and time,
only 50 matrices of size 200 $\times$ 200 have been analyzed. 
The level spacing statistics for the generated matrices is compared to 
the analytical result. 

Before attempting to generate large normal matrices the algorithm was
checked with $2 \times 2$ matrices. As these were easily produced,
11,000 were generated and compared to the exact analytical
result,
\begin{eqnarray}
p_{2\times 2}(s) &=& {{\pi^2s^3}\over 10.4683}e^{-{1\over 4}s^2}
I_0({s^2\over 4}).
\end{eqnarray}

\begin{center}
\begin{figure}[bh]
\epsfig{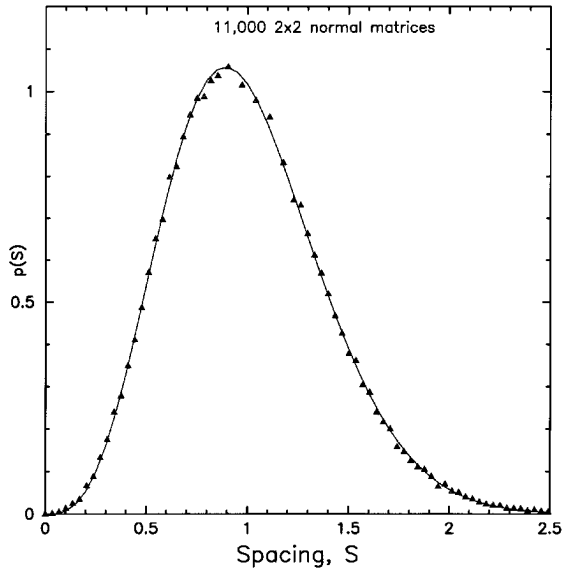}
\caption{The level spacing distributions, p(S), for $N=2$, 
analytical form and 11,000 numerically generated normal matrices.} 
\label{fig:num5}
\end{figure}
\end{center}

The eigenvalues are unfolded so that the mean spacing is unity, this
is achieved by rescaling the spacing and renormalizing the spacing
distribution so that the mean spacing between eigenvalues is unity,
\begin{eqnarray}
S & \equiv & 1.6622 s, \nonumber \\
p(S) & \equiv & p(s)/10.4883.
\end{eqnarray}
Plots of the analytical form and the unfolded eigenvalues are displayed in
figure \ref{fig:num5}.

\begin{center}
\begin{figure}[bh]
\epsfig{file=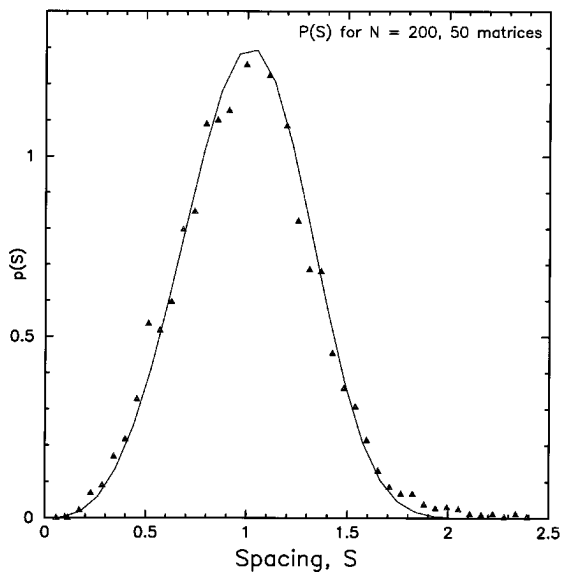}
\caption{
The level spacing distributions, p(S), for $N=200$, analytical result 
(with eigenvalues unfolded), and 50 numerically generated $200\times 200$
normal matrices.}
\label{fig:num6}
\end{figure}
\end{center}

For normal matrices of size $200\times 200$ the exact analytical form
for the spacing distribution can easily be found from equations
(\ref{esw})-(\ref{psw}),
\begin{eqnarray}
p_{200} (s) &=& (199-\sum_{q=1}^{199} {e_{q-1}(s^2)\over e_q(s)})
2 s \prod_{i=1}^{199} \left[ e^{-s^2} e_i (s^2)\right] , \label{ps200}
\end{eqnarray}

where $e_q(s^2) = \sum_{l=0}^{q} {s^{2l}\over l!}$.
Under the optimization algorithm it is found that many eigenvalues
``leak'' out into the complex plane. Adding a further constraint to 
keep the eigenvalues within a radius of $\sqrt{2\sigma N-1} \approx 20$
lengthens the generation time dramatically. 
Therefore, for this preliminary study the
eigenvalues were ``pruned'', those falling outside $r=18.1$ were
neglected, (the number was choosen so that edge effects would be 
eliminated as well). Overall, about one third of the eigenvalues were 
pruned. The results are displayed in figure \ref{fig:num6}.

The slight deviation of the numerical and analytical results for 
large $S$ is due to the leakage of eigenvalues. The small $S$
behavior is not affected by the leakage. It is found that changing
the initial conditions has little effect on $p(S)$. It is also 
observed that lowering the the threshold for the error function by
a factor of ten has a negligible effect.

\section{Conclusion}

It has been shown that cubic repulsion of eigenvalues obtained
from random normal matrices experience a minimum of cubic 
repulsion in the complex plane. Even though normal matrices
are subsets of complex matrices we have demonstrated that 
Gaussian normal matrix models yield the same results as the Ginibre
ensemble of complex matrices. One can define ensembles of complex
matrices with weights invariant under similiarity transformations,
(see appendix B of \cite{oas}). These ensembles will fall under the
$\beta =4$ and experience a minimum of quartic repulsion. (Getting
analytical results for the statistics of the eigenvalues is 
difficult however). 
Ensembles of complex matrices with weight functions invariant under
unitary transformations other than the Gaussian are not factorable
and are not amenable to analytical analysis. For normal matrices we
can define identical ensembles to complex models invariant under
similiarity transformations and as well define a wide range of
ensembles invariant under unitary transformations, (as demonstrated
in this paper).
The special case of real normal matrices
was also shown to experience cubic repulsion in the complex plane. 
As well,
there are even more significant differences between random normal
matrices and, more traditional, random matrix models composed
of hermitian, unitary, and real symmetric matrices, themselves being
subsets of normal matrices. 
Although numerical generation of normal matrices is not as
straightforward as for other types of matrices we have shown that analytical 
results are often much easier to obtain. What numerical 
information we could obtain does agree with the theoretical result.

It is a little disappointing that universality, in the sense of 
Brezin and Zee \cite{brezinzee}, is not found for normal matrix
models. The simple scaling arguments
that lead to universal kernels for other types of matrices do 
not work here.
It may be possible to rescale $S$ to obtain a universal
spacing distribution $P(S)$. 
This allows the
potential of analytically exploring the statistics of complex
eigenvalues which was not possible before. Another interesting
feature is the closed form expressions for the two point 
correlation functions in NGE, Laguerre ($a=0$), and Legendre
ensembles in the infinite matrix limit, (see Table \ref{table}). 

In terms of physical applications it should be noted that results
using the Ginibre ensemble of complex matrices can be duplicated
with the normal Gaussian ensemble. Such work includes,
modeling cellular structures by Le Caer and Ho \cite{voronoi},
two dimensional quantum gravity by Morris \cite{morris}, and
Haake's work on generators of quantum dynamics for systems with
dissipation \cite{haake}. 
It has recently been shown, \cite{oas2}, that Haake's generator of dissipation,
$D$, in \cite{haake2} \cite{haake}, is properly represented by
a normal matrix rather than a complex one. This result strengthens
the idea that the eigenvalues of these generators universally exhibit cubic
repulsion when dissipative dynamics is present. 

The results obtained in this work give a more complete picture of the 
statistics of eigenvalues from random matrices. The normal matrix
model, a link between the Dyson ensembles and the Ginibre ensemble,
has a rich structure where interesting results can still be found.


\widetext
\begin{table}

\caption{Ensembles of random normal matrices defined by their
probability distribution. Also shown are the corresponding
eigenvalue density and two point function. For the NGE, Laguerre $a=0$,
and Legendre ensembles the closed form large $N$ result has been shown.}
\label{table}
\begin{tabular}{llll} 
Ensemble & Probability Distribution & Density, $\rho$ & $\rho (r_1)\rho (r_2)-
R_2$ \\ \tableline
{\small NGE} &{ $ e^{-Tr{\bf  M^{\dag} M}}$} &{ ${1\over \pi }$} &{ $ {1\over \pi ^2}\exp [-|z-w|^2]$}   \\ 
{\small NQE}  &{ $e^{-Tr({\bf M^{\dag} M}-g{\bf M^{\dag} M^{\dag}  MM})}$} &
{ $ {e^{-r^2-gr^4}\over
\pi}\sum_{l=0}^{N-1}{r^{2l}\over N_l(g)}$} & 
{ ${e^{-r_1^2-r_2^2-gr_1^4r_2^4}\over \pi ^2}K_N(z_1,z_2)K_N(z_2,z_1)$}
\\  
{\small Laguerre} & { $e^{-Tr (({\bf M^{\dag} M})^{1\over 2}+{a\over 2}\ln
({\bf M^{\dag} M}))}$}&{ ${e^{-r}r^{a+1}\over 2\pi}\sum_{l=0}^{N-1}
{r^{2l}\over \Gamma [2l+a+2]}$} & 
{$ \sum_{l=0}^{N-1}{(r_1r_2)^{a+1}e^{-r_1-r_2} z_1^l{z_2 ^{\ast}}^l{z^{\ast}}_1^m
z_2^m\over
4\pi ^2 \Gamma [2l+a+2] \Gamma [2l+a+2]}$} \\ 
{\small Laguerre, $a=0$} & {$ e^{-Tr ({\bf M^{\dag} M})^{1\over 2})}$} 
&{ ${e^{-r}\over 2 \pi r}\sinh (r)$} &{ $ 
  {e^{-r_1-r_2}  \over 4\pi ^2 r_1 r_2}
\sinh (\sqrt{z_1z_2 ^{\ast}})\sinh(\sqrt{z_1 ^{\ast} z_2})
$ } \\ 
{\small Legendre} &{ 1, (evals $\in$ unit circle)}	&
 ${1\over \pi (1-r^2)^2}$ & ${1\over \pi ^2}\left[ 
{1\over (1-z_1 z_2 ^{\ast})}^2{(1-z_1 ^{\ast} z_2)}^2\right] $ \\  
{\small Jacobi}   & $det[1-{({\bf M^{\dag} M})}^{1\over 2}]^a$	
&{${(1-r)^a\over 2\pi }
\sum_{l=0}^{N-1} {r^{2l}\over B[a+1,2l+2]}$ }
&
{ ${1\over 4\pi ^2}\sum_{l=0}^{N-1} {{z_1} ^l {z_2^{\ast}}^l {z^{\ast} _1}^m z_2
^m\over 
B[a+1,2l+2] B[a+1,2m+2]}$} \\ 
{\small Gen. Gaussian} &
{ $e^{-Tr({\bf M^{\dag} M}-{a\over 2}\ln ({\bf
M^{\dag} M}))}$} &{ ${e^{-r^2}r^a\over \pi}\sum_{l=0}^{N-1} {r^{2l}\over 
\Gamma [l+{a\over 2}+1]}$} &
{ $\sum_{l=0}^{N-1}{(r_1r_2)^a e^{-r_1 ^2 r_2 ^2} {z_1} ^l{z_2^{\ast}}^l {z^{\ast} _1}^m z_2 ^m\over \pi ^2 \Gamma [l+{a\over
2}+1]\Gamma [m+{a\over 2}+1]}$ } \\  
 
\end{tabular}

\end{table}

\end{document}